\documentclass[aps,prl,twocolumn,
showpacs,preprintnumbers,amsmath,amssym,superscriptaddress]{revtex4}
\usepackage{graphicx}
\usepackage{bm}
\usepackage{amssymb}
\begin{document} 

\title{
Effective Langevin equations for the pair contact process with diffusion}

\author{Ivan Dornic}
\affiliation{\mbox{CEA -- Service de Physique de l'\'Etat Condens\'e,
~Centre~d'
\'Etudes~de~Saclay,~91191~Gif-sur-Yvette,~France}}

\author{Hugues Chat\'e}
\affiliation{\mbox{CEA -- Service de Physique de l'\'Etat Condens\'e,
~Centre~d'
\'Etudes~de~Saclay,~91191~Gif-sur-Yvette,~France}}

\author{Miguel A. Mu\~noz}
\affiliation{\mbox{Instituto~de~F\'\i sica~Te\'orica~y
~Computacional~Carlos~I,
~Universidad~de~Granada,~18071~Granada,~Spain}}

\date{\today}

\begin{abstract}
We propose a system of coupled, real-valued, effective Langevin
equations for the nonequilibrium phase transition exhibited by the
pair contact process with diffusion (and similar triplet and
quadruplet, $n$-uplet, processes). A combination of analytical and
numerical results demonstrate that these equations account for all
known phenomenology in all physical dimensions, including estimates of
critical exponents in agreement with those reported for the
best-behaved microscopic models. We show in particular that the upper
critical dimension of these $n$-uplet transitions is $\frac{4}{n}$,
and $\frac{4}{n} -1$ for their anisotropic (biased) versions.
\end{abstract}

\maketitle
Whereas there exists ample evidence of universality among
out-of-equilibrium phase transitions, their full classification is by
no means achieved. At equilibrium, coarse-grained descriptions in
terms of Ginzburg-Landau-Wilson free-energy functionals and their
associated Langevin equations encode in a systematic way the
symmetries, conservation laws, and dimensionalities determining
universality classes. Out of equilibrium, even the relevant
ingredients are still debated, and continuous stochastic descriptions
are often missing, so that the non-equilibrium counterpart of the
celebrated taxonomy by Hohenberg and Halperin \cite{HH} remains a
distant achievement. Here we are interested in irreversible phase
transitions into fluctuation-less absorbing states.  Such transitions
abound in non-equilibrium physical phenomena such as epidemics,
catalysis, synchronization, or self-organized criticality
\cite{Reviews_AS}. The prominent directed percolation (DP) class,
which encompasses all phase transitions into a single absorbing state
(without additional symmetries, conservation laws, or disorder),
well-established both theoretically and numerically, can be
characterized by the following Langevin equation for a single
coarse-grained, real density field $\rho=\rho(x,t)$:
\begin{equation}
\label{sde_DP}
{\partial_t}\rho=D\, \nabla^2\rho +a \rho-b \rho^2 +\sigma
\,\rho^{\frac{1}{2}}\,\eta
\end{equation}
where $\eta=\eta(x,t)$ is a Gaussian white noise whose influence vanishes
in the absorbing state ($\rho \equiv 0$).  In spite of some recent progress
in the search for Langevin descriptions of other classes of absorbing
phase transitions, the situation of the DP class remains
exceptional. Indeed, even if Eq.~(\ref{sde_DP}) can be derived
rigorously %using well-trodden techniques
for some reaction-diffusion models in this class, 
such Langevin equations sometimes simply do not exist.
%many other cases lead to field theories that correspond to Langevin equations which are either very hard to renormalize, or which sometimes simply do not exist. 
For example, for the
simple annihilation process $2A \to 0$ it is well-known that one
obtains ``imaginary noise'' (see below), which precludes a
well-behaved Langevin equation \cite{Nature}.

Reaction-diffusion systems where creation and annihilation processes
require pairs of particles (while isolated ones can diffuse but not
react) are usually considered to represent the {\it pair-contact
process with diffusion} (PCPD) class. Studies of various models in
this (putative) class have produced a series of conflicting results
and opinions (see \cite{Review_PCPD} for a review and
references therein). From the numerical side it is still
debated whether in $d=1$ the PCPD behaves as DP or not, as numerics
are often plagued with long transients
\cite{Review_PCPD,Haye_cyclic,Haye_CA,Kock03,Barkema,Haye05}.
What seems to be widely accepted is that the upper critical dimension
$d_{\rm c}$ is not $4$ as in DP, but $d_{\rm c}=2$
\cite{OdorPorto,Odor2d,Review_PCPD}. 
 It has also been shown numerically that when
a bias (anisotropy) is introduced in the diffusion of isolated
particles, the critical dimension is reduced to $d_{\rm c}=1$ (at odds
with what happens for DP where biased diffusion plays no significant
role)
\cite{Korean_drift}.  
At the analytical level Janssen {\it et al.} \cite{Fredtrouduc04}
recently showed that a perturbative renormalization group analysis of
the field theory derived from the microscopic reactions $2A \to 3A, 2A
\to \emptyset/A$ (together with diffusion of isolated $A$ particles) yields
only unphysical fixed points and runaway trajectories.
Generalizations of the PCPD requiring triplets or quadruplets of
particles for the reactions (respectively called TCPD and QCPD for
``triplet and quadruplet contact process with diffusion''; $n$-CDP in
general) have also been studied, with results at least
as controversial as those for the PCPD \cite{Haye_Korean,Odor_34,Kock03}.

What makes these problems interesting beyond the usual skirmishes
between specialists is that the ingredient which seems to lead out of
the DP class is the fact that $2$ ($n$, in general) particles are
needed for reactions to occur. That such a microscopic constraint, not
related to any new symmetry nor conservation law, should determine
universal properties may indeed appear unacceptable, hence the
opinion, held by some authors \cite{Haye_CA,Barkema,Haye05},
 that the observed
critical behavior will eventually turn up to be in the DP class.
Elucidating the critical behavior of the PCPD
has thus become a milestone in the ongoing debate about universality
out-of-equilibrium.

In this Letter, we propose a system of coupled, real-valued, effective
Langevin equations for the nonequilibrium phase transition exhibited
by the PCPD (and of the allied $n$-CPD models).
 Analyses of these equations allow us: i) to show that
$d_{\rm c}=\frac{4}{n}$ with a reduction by unity when a biased
diffusion is switched on, ii) to reproduce the correct (mean-field)
critical exponents above $d_{\rm c}$ in all cases, and iii) to
estimate by direct numerical integration of these equations critical
exponents in agreement with those of the best microscopic models in
any dimension.

A continuous description of the PCPD problem, if possible at all, should be
in terms of at least two fields as converging conclusions signal
\cite{Korean_drift,Fredtrouduc04}. Indeed, a two-field Langevin
description has been successfully proposed for the related pair
contact process {\it without} diffusion, the prototypical model with
infinitely-many absorbing states \cite{PCP}.  As for that case, here a
``pair-field'' and a ``singlet field'' is the most natural choice.
To build up a set of Langevin equations, we start from the following
microscopic two-species reaction-diffusion model in the PCPD family
\cite{Haye_cyclic}: pair-activity is represented by usual DP-like
reactions $B \to 2B, B \to 0$ while isolated particles diffuse and
undergo pair annihilation $2A
\to \emptyset$. The two species are coupled cyclically: $B\to 2A$ (a
``pair'' separates into $2$ particles) and $2A\to B$ ($2$ particles
recombine into a ``pair''). 
%\cite{NOTEDIF}.
 Using well-trodden techniques
(see, e.g., \cite{Fock}), one arrives at  the following coupled
equations: 
\begin{eqnarray}
\label{eq-psi}
\partial_t \psi_B &=& D_B \nabla^2 \psi_B + a_B \psi_B -b_B \psi^2_B
+c_B \psi^2_A
%+ d_B \psi_A \psi_B  
+ \sigma_B \psi^{\frac{1}{2}}_B \eta_B
\nonumber \\
\partial_t \psi_A &=& D_A \nabla^2 \psi_A -b_A \psi^2_A
+c_A \psi_B
%-d_A \psi_A \psi_B 
+ i \sigma_A \psi_A \eta_A,
\end{eqnarray}
where all coefficients depend on microscopic reaction rates and
diffusion constants, and $\eta_A$ and $\eta_B$ are Gaussian white
noises (including some cross correlations not specified here).
The mapping through the two-species model disentangles 
 the competition between  the real, DP-like component of the noise,
and the imaginary one (arising  from the annihilation reaction
$2A\to\emptyset$) \cite{Howard_Tauber}.
But now both fields  $\psi_A$ and $\psi_B$ are complex,
and  even though their noise-averages   do correspond to the local densities
$\rho_A$ and $\rho_B$ of $A$ and $B$ particles,
the same is not true for  %their
 higher-order moments, impeding physical intuition.
%And  a renormalization group  analysis of these equations appears to be 
%a  bewildering complicated task.

\begin{figure}
\includegraphics[width=8.0cm,clip]{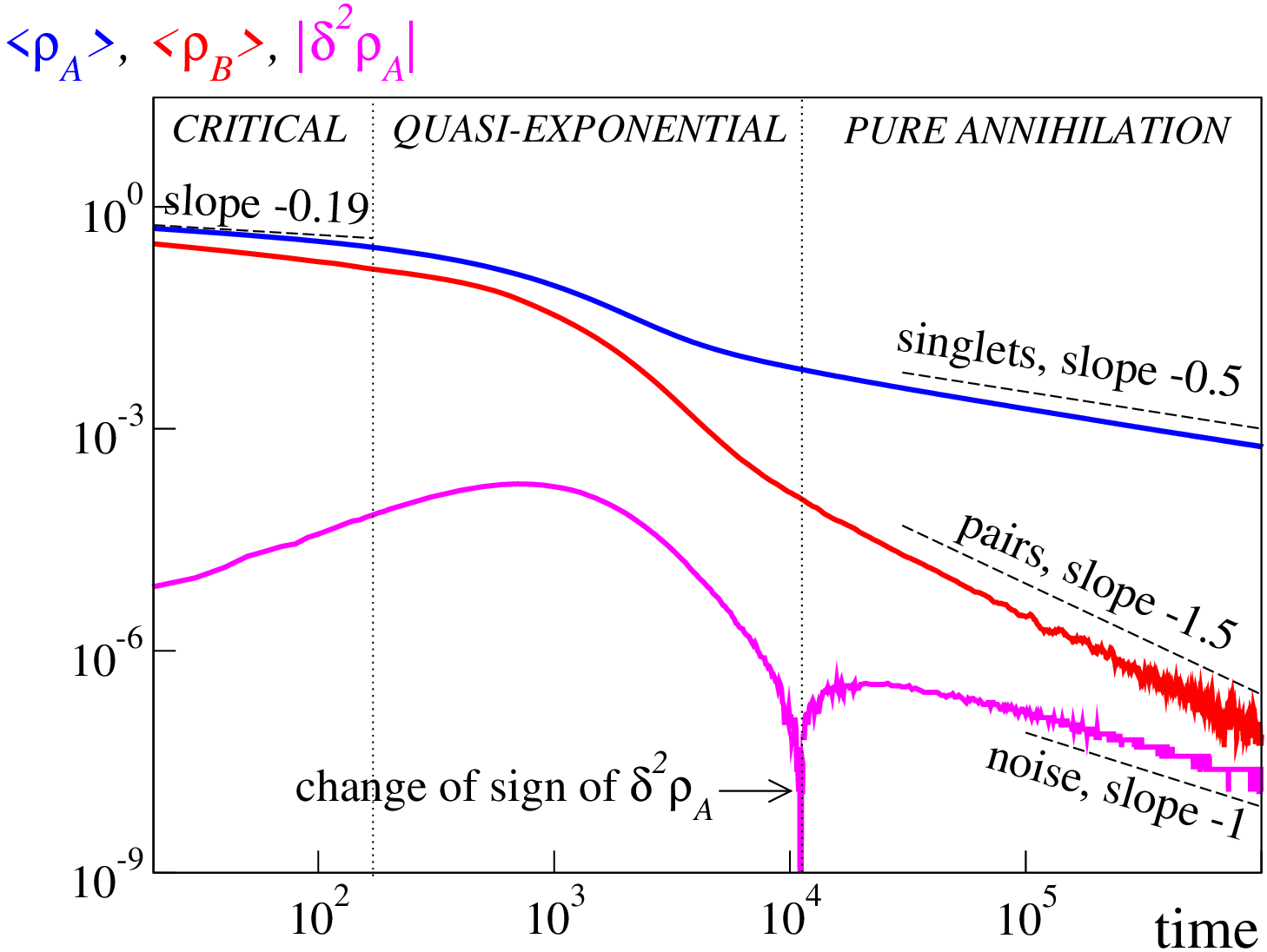}
\caption{(Color online) 
Evolution, during a slightly subcritical quench, of the densities of
isolated particles $A$ and pair particles $B$, together with the
absolute value of the singlet-field noise variance $(\delta\rho_A)^2$
measured as the connected correlation function of the local density of
$A$s calculated between neighboring blocks of size $64$.  Two-species
PCPD model described in text implemented as in
\protect{\cite{Kock03}} ($d=1$, $2^{23}$ sites, $16$ runs).  }
\label{fig1}
\end{figure}

However, perusal of simulation results of PCPD-like models suggests that {\it
annihilation is only predominant deep inside the absorbing phase, and
not in the active one nor around the critical point}.  Indeed,
slightly subcritical quenches reveal a quasi-exponential departure
from the critical power-law scaling, with the typical annihilation
power-law decay ($t^{-1/2}$ for singlets and $t^{-3/2}$ for pairs)
setting up only at a much later time (see  Fig.~\ref{fig1}
and also \cite{Haye_CA}).
 Correspondingly, the variances of the noise terms  
---which can be estimated from
measurements of block-spin like correlations of particle densities in
microscopic models--- are initially DP-like (i.e. positive and
proportional to the particle densities) in the near-critical decay regime,
 then  change sign (around
$t=10^4$ in  Fig.~\ref{fig1}; see also \cite{Haye_wetting}), 
and finally become
negative (imaginary noise).

%%%%%%%%%%%%%%%%%%%%%%%%%%%%%%%%%%%%%%%%%%%%%%%%%%%%%%%%%%%%%%%%%%%%

On the basis of the above observations, 
and since our aim is to build an effective Langevin description of PCPD
{\it critical points},
%(and not of the well understood annihilation-dominated absorbing phase),  
we simply discard the imaginary 
noise term (and its associated anticorrelations), and
modify the nonlinear terms for the singlet-field so that
it decays, for the space dimension of interest and when the pair-field is set 
to zero, with the appropriate decay at least on average \cite{Redner}.
The abstract fields $\psi_A,\psi_B$  in Eqs.(\ref{eq-psi}) then
become the real particle density fields $\rho_A$ and $\rho_B$, 
governed by the following effective Langevin equations:
\begin{eqnarray}
\label{pcpd-sde}
\partial_t \rho_B &=& D_B \nabla^2 \rho_B + a \rho_B -b \rho^2_B
+c \rho^\alpha_A + \sigma_B \rho^{\frac{1}{2}}_B \eta
\nonumber \\
\partial_t \rho_A &=& D_A \nabla^2 \rho_A -e \rho^\alpha_A +f \rho_B, 
\end{eqnarray}
which themselves change with dimension:  for $d=1$, one has $\alpha=3$ since the singlet-field must
decay like $1/\sqrt{t}$ for pure annihilation, while
for $d>2$  the mean-field pure-annihilation decay
$1/t$ imposes $\alpha=2$. In the marginal case $d=2$, the
annihilation decay is $\ln t/t$, and  the incriminated terms in
Eqs.(\ref{pcpd-sde}) have to be completed by logarithmic factors
($\pm\rho^2_A/\ln{(1/\rho_A)}$). 
From these ``bare'' equations, one  readily realizes that some extra terms 
---in particular a cross term  $\omega \rho_A \rho_B$ in the $\rho_B$ equation
 or a noise term $ \sigma'  \sqrt{\rho_B} \eta_A$ 
in the $\rho_A$ equation--- are generated 
 perturbatively once the effect of fluctuations is
 taken into account.
Thus these terms should be incorporated from the beginning, %/onset
and disregarding now all noises and  derivatives, one 
arrives at  the mean-field decay exponents
 $\theta_A^{\rm MF}=\frac{1}{\alpha}$ and
$\theta_B^{\rm MF}=1$, in agreement with the known results for PCPD
(where $\alpha=2$ in high dimensions) \cite{Review_PCPD}.

\begin{figure}
\includegraphics[width=8.0cm,clip]{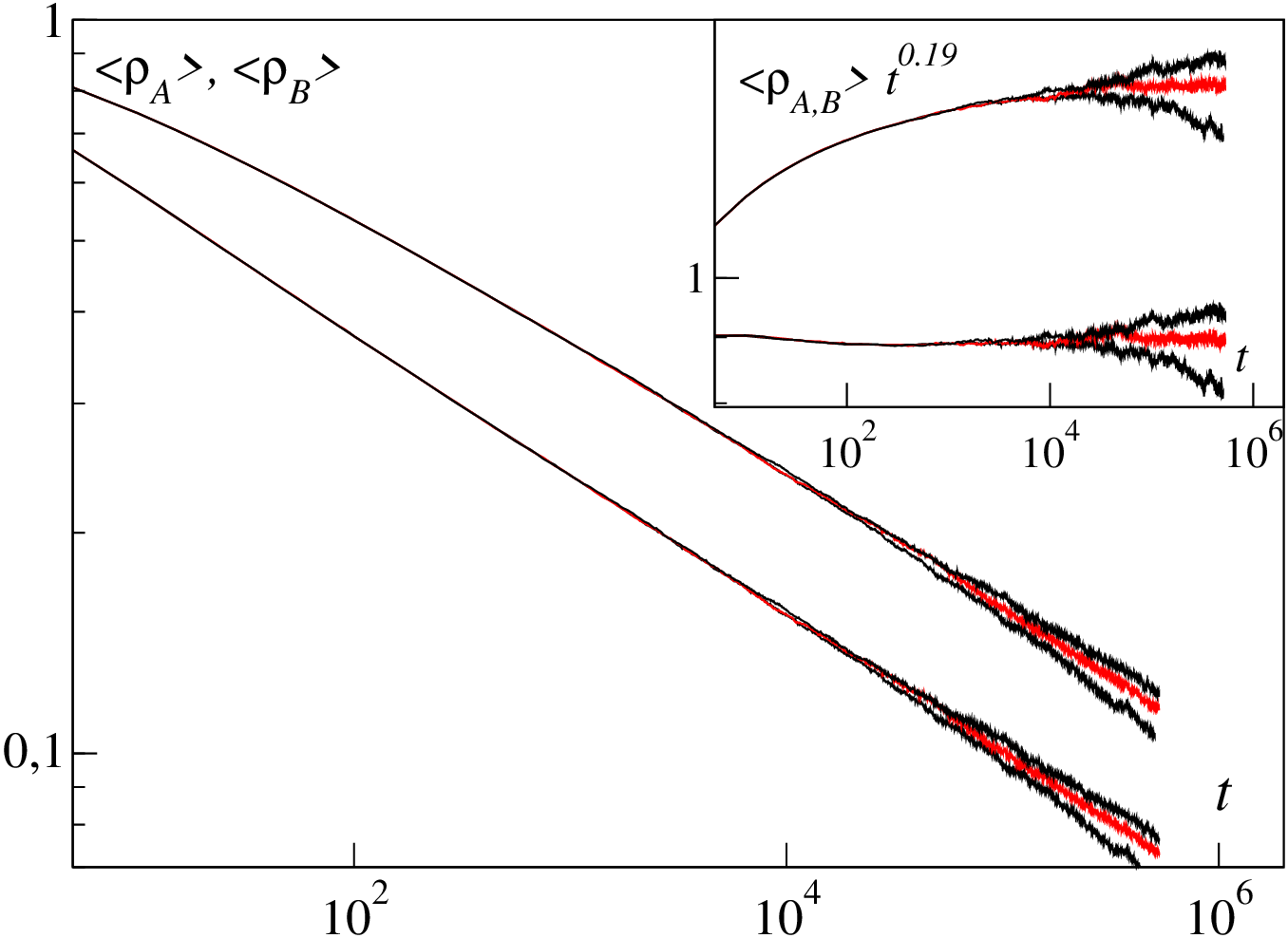}
\caption{
(Color online) Numerical integration of Eqs.(\ref{pcpd-sde}) in $d=1$
($D_A=0$, $D_B=0.25$, $b=c=e=f=1$, $\sigma_B^2=2$, $\Delta x=1$,
$\Delta t=0.25$, single run on system of size $2^{23}$).  Decay of
the (spatially-averaged) densities
$\langle\rho_A\rangle$ (top) and $\langle\rho_B\rangle$ (bottom)
during near-critical quenches from initial condition $\rho_A=0$,
$\rho_B=1$ for $a=1.4442$, $1.4443$ and $1.4444$. Inset: same
multiplied by $t^{0.19}$.}
\label{fig2}
\end{figure}

After having verified the validity of our Langevin equations in high
dimensions (mean-field) we now study their behavior in low dimensions
by direct numerical integration. For simplicity we first integrate
Eqs.(\ref{pcpd-sde}), and afterwards check that the extra
perturbatively generated terms do not alter the critical behavior.
Note that the equation for the pair-field $\rho_B$ is almost identical
to the DP one (\ref{sde_DP}) and, therefore, the accurate and
efficient integration scheme recently presented in \cite{Us1} can be
employed. Varying the ``temperature-like'' parameter $a$, an absorbing
phase transition is observed, with (apparently) clean scaling behavior
at and around the critical point in all tested physical dimensions.
As expected from our model-building,
 subcritical quenches show a quasi-exponential departure
from the scaling regime, followed by a late annihilation-like decay
({\it without} local anti-correlations).  In $d=1$ critical quenches
allow to estimate the decay exponent
$\theta=\beta/\nu_\parallel=0.19(1)$ (Fig.~\ref{fig2}). Measuring
steady-state activity above threshold  determines
$\beta=0.33(2)$, while finite-size lifetime statistics at criticality
or seed simulations yield indifferently
$z=\nu_\perp/\nu_\parallel=1.64(3)$ (not shown). Even though the
quality of these results, at par with those of the ``best''
microscopic models, leads to suggest asymptotic values different from
that of the DP class in $d=1$, we prefer in view of recent large-scale
simulations in PCPD problems \cite{Haye05}, to avoid drawing a
conclusion on whether the ``true'' asymptotic behavior is DP-like or
not, based solely on such a small quantitative difference (for the DP
class $\theta\simeq 0.16$, $\beta\simeq 0.28$, and $z\simeq 1.58$.)
We have extensively checked that these results are robust against both
changes in parameter values and inclusion of the extra perturbatively
generated terms.

\begin{figure}
\includegraphics[width=8cm,clip]{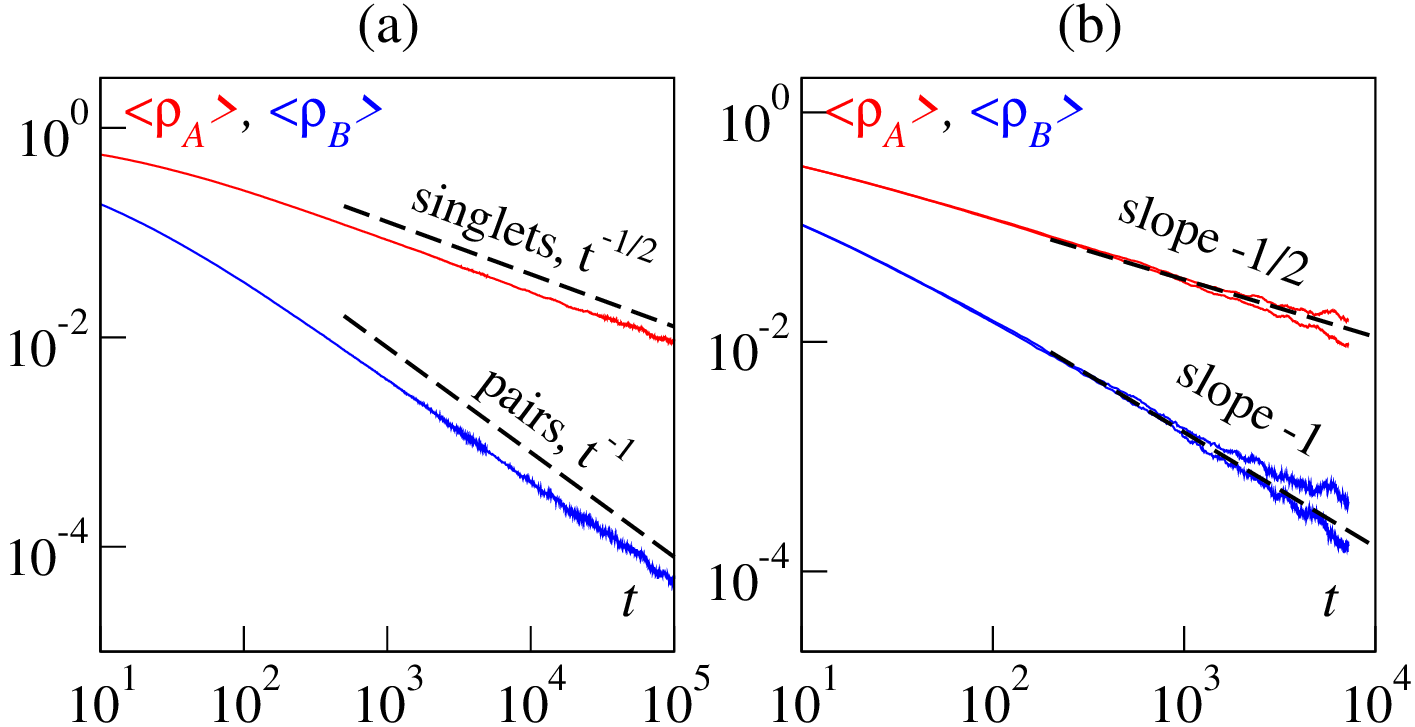}
\caption{
(Color online) Decay of singlet and pair densities during
near-critical quenches for $d=3$ PCPD models. (a) Reaction-diffusion
model $2A\to3A$, $2A\to\emptyset$ implemented as in
\protect{\cite{Kock03}}; (b) Integration of the Langevin
Eqs.~(\ref{pcpd-sde}) with $\alpha=2$.}
\label{fig3}
\end{figure}

In higher dimensions, the difference with DP transitions becomes
qualitative: our simulations in $d=2$ show that at threshold both
fields decay with the same exponent $\theta^{d=2}\simeq 0.50(2)$, the
remaining curvature in log-log plots (not shown) indicating the
presence of logarithmic behavior for the ratio $\langle
\rho_B \rangle/\langle \rho_A \rangle$ characteristic of the marginal
 dimension (in agreement with microscopic models). For $d=3$,
 simulations of Eqs.(\ref{pcpd-sde}) with $\alpha=2$ show different
 scaling at threshold for the two fields: $\theta_A^{d=3}\simeq
 0.50(5)$ and $\theta_B^{d=3}\simeq 1.0(1)$, values according with
 those found in microscopic models (Fig.~\ref{fig3}), and with the
 previously found mean-field predictions. This, and the behavior in
 $d=2$ suggests that $d_{\rm c}=2$.

We now turn our attention to power-counting arguments aimed at
identifying the critical dimension analytically and determining the
relevance or irrelevance of the different terms of our Langevin
equations. As usual in problems with multiple fields, one has some
freedom to perform naive scaling analyses. If we enforce, as usual,
the coefficient of $\partial_t \rho_B$ to scale as a constant (fixing
in this way the dominant time-scale) and the associated action to be
dimensionless, one is ineluctably led to the conclusion that the noise
term in the $\rho_B$-equation is the dominant nonlinearity together
with the saturation term $-b \rho_B^2$, and that both become relevant
below $d_{\rm c}=4$ as in DP. This being in contradiction with our
findings, it is mandatory to scale in a different way. If we now
impose the coefficient of the other time-derivative,
$\partial_t\rho_A$, to scale as a constant, we conclude that the main
nonlinearity is the (generated) noise term of the $\rho_A$ equation,
proportional to $\sqrt{\rho_B}$, which becomes relevant below $d_{\rm
c}=4/\alpha$. Therefore $d^{\rm PCPD}_c=2$ (as $\alpha=2$ in PCPD above
 its critical dimension), in agreement with our numerics.
 One can thus consider that the (activity) pair-field $\rho_B$ is ``slaved'' to
the singlet field $\rho_A$ (which is the leading one, controlling the
scaling). This viewpoint is further supported by the numerical
observation that the decay-exponent $\theta_B$ experiences a
discontinuity at $d_{\rm c}$ where it jumps from its mean-field value
$\theta_B=1$ above $d_{\rm c}$ to $\theta_B=\theta_A$ at or below. Following
the above discussion, our final minimal set of effective equations
yielding the correct scaling laws (as we have verified numerically)
both when fluctuations are relevant and when they can be neglected
reads:
\begin{eqnarray}
\label{pcpd-final}
\partial_t \rho_B &=&  D_B \nabla^2 \rho_B + a \rho_B -b \rho^2_B +
c \rho^\alpha_A +\omega \rho_A\rho_B
\nonumber \\
\partial_t \rho_A &=& D_A \nabla^2 \rho_A -e \rho^\alpha_A +f \rho_B 
+ \sigma_A \rho^{\frac{1}{2}}_B \eta.
\end{eqnarray}
Note the change of perspective from
Eqs.(\ref{pcpd-sde}), which leads to consider the PCPD (and related models) 
as a system of annihilating random walks wandering
between high-activity (pair) clusters, themselves of course 
self-consistently determined by the complex interplay of the 
two modes of the dynamics.
These clusters  act not only  as non-trivial  boundaries for 
the isolated particles,  but  also as   fluctuating
  {\it sources}.
  Since fluctuation effects for $2A \to \emptyset$ (or $2A \to A$) 
in the presence of  even a constant source term
are known to be relevant up to and 
{\it including} $d=2$ \cite{RaczDroz}, this provides, in our opinion,
an intuitive  mechanism which can explain, among other things,
 why $d_c=2$ for the PCPD and
also why the dynamical exponent is not that of simple random walks
\cite{NOTEBOUNDARY}.

The two-species/two-fields description put forward above is easily
extended to the TCPD and QCPD cases \cite{Haye_Korean,Odor_34,Kock03},
by replacing the pair field by a triplet or quadruplet field: the
interpretation of the fields changes but the governing equations
(\ref{pcpd-final}) remain identical, with an exponent $\alpha$ chosen
according to the corresponding behavior of the pure annihilation
processes ($3A\to\emptyset$ or $4A\to\emptyset$).
For $3A\to\emptyset$, the particle density decays as $\sqrt{\ln t/t}$
in $d=1$, which can be accounted for by $\alpha=3$
 with a logarithmic correction,
while for $d>1$ mean-field decay $1/\sqrt{t}$ (corresponding to
$\alpha=3$) emerges. A striking consequence is that  the TCPD class 
should not exist {\it per se}
in $d=1$, since its effective Langevin equations are identical to
those of the PCPD class (the logarithmic term plays no r\^ole on
asymptotic scaling). Moreover, following the power-counting analysis
sketched above, $d_{\rm c}^{\rm TCPD}= \frac{4}{3}$ (in agreement
with \cite{Haye_Korean}), with the
mean-field exponents $\theta_A=\frac{1}{3}$ and $\theta_B=1$. Now for
quadruplet models, the pure-annihilation mean-field decay $t^{-\frac{1}{3}}$
holds  in
all physical dimensions and thus $\alpha=4$ always,
leading to $d_{\rm c}^{\rm QCPD}=1$. Indeed, in $d=1$ both fields
decay with exponent $\theta=\frac{1}{4}$ with 
strong log corrections (most likely 
$\langle \rho_A \rangle/\langle \rho_B \rangle
\sim (\ln{t})^{1/3}$), while
for $d>1$, $\theta_A=\frac{1}{4}$ and $\theta_B=1$ (mean-field
results). All these results (in disagreement with some previously
published ones) have been verified by simulations of the Langevin
equations and of microscopic two-species reaction-diffusion models,
and will be reported elsewhere.

Finally, let us discuss the r\^ole of biased diffusion (anisotropy) for
the isolated particles within our picture \cite{Korean_drift}. 
Starting either directly from microscopic models, or from symmetry
considerations,  one realizes that an additional $\nabla \rho_A$-term needs to
be introduced in the $\rho_A$-equation. A %naive
 power counting analysis, analogous to the one
discussed above, then reveals that $d_c$ is reduced by one unit, so $d_c=1$
for $\alpha=2$ (biased-PCPD) and it is below $d=1$ for larger values
of $\alpha$: the biased TCPD and QCPD are expected to exhibit
mean-field behavior in any physical dimension. All these results are
in agreement with known ones \cite{Korean_drift}
 and have been verified numerically.

Summing up, we have proposed effective coupled Langevin equations
governing the densities of a $n$-uplet field and a singlet field in
reaction-diffusion problems where at least $n$ particles are required
for creation or annihilation to occur. A combination of numerical and
analytical arguments have shown that they do reproduce the behavior of
microscopic models, and thus provide a sound framework to further
study these problems. Such future work should concentrate on a
renormalization group treatment of these equations, possibly in a
non-perturbative approach.  Due to the closeness of estimated exponent
values for the PCPD case in $d=1$ with those of the DP class, this
appears as the only solution to put an end to the crucial question of
whether such problems, below their upper critical dimension, exhibit
genuinely novel critical behavior or not.

M.~A.~M. acknowledges financial support from the Spanish MCyT (FEDER)
under project BFM2001-2841.

\end{document}